\documentclass[aps,prl,twocolumn,superscriptaddress,showpacs]{revtex4}
\usepackage{amsfonts,amscd,amsmath, amssymb,graphicx,color}

\def\ep{\text{e}}

\def\g{\mathfrak{g}}
\def\s{\mathfrak{s}}

\def\4{\tfrac{1}{4}}
\def\nq{\text{\tiny NQ}}
\def\n{\text{\tiny N}}
\def\qqb{\text{\tiny Q}\bar{\text{\tiny Q}}}
\def\3q{\text{\tiny 3Q}}

\begin{document}
\preprint{LMU-ASC 24/15}
\title{Model of the $N$-Quark Potential in $SU(N)$ Gauge Theory using Gauge-String Duality}
\author{Oleg Andreev}
\affiliation{L.D. Landau Institute for Theoretical Physics, Kosygina 2, 119334 Moscow, Russia}
\affiliation{Arnold Sommerfeld Center for Theoretical Physics, LMU-M\"unchen, Theresienstrasse 37, 80333 M\"unchen, Germany}
\begin{abstract}
We use gauge-string duality to model the $N$-quark potential in pure Yang-Mills theories. For $SU(3)$, the result agrees remarkably well with lattice simulations. 
The model smoothly interpolates between almost the $\Delta$-law at short distances and the Y-law at long distances. 
\end{abstract}
\maketitle

\textit{Introduction.---}
Predicting properties of hadrons still represents a serious challenge for Quantum Chromodynamics (QCD). Heavy quarks closely resemble static test charges and therefore are useful to probe confining properties of QCD. So far, great progress has been made in the study of quarkonia, i.e. mesonic states that contain two heavy constituent quarks. In contrast, systems of three or more heavy quarks, which are a good starting point for understanding the phenomenology of baryons and multi-quark bound states, are much less studied. In this case a key issue is whether multi-quark interactions can be understood in terms of two-body interactions or whether there are genuine three- and many-body effects to be considered as part of the overall picture of strong interactions \cite{bj,richard}.

The best known phenomenological models of the $N$-quark potential are those of $N=3$, the so called $\Delta$ and Y-laws \cite{review}. The $\Delta$-law is based on pairwise interactions between quarks, while the $Y$-law is an examples of three-body interactions. In the infrared region the former predicts that the potential grows linearly with the perimeter of the triangle formed by quarks \cite{delta-law}, while the latter predicts a linear growth with the minimal length of a string network which has a junction at the Fermat point of the triangle \cite{Y-law}. 

Until recently, lattice gauge theory was the premier method for obtaining quantitative and qualitative information about strongly interacting gauge theories. For the three-quark potential the accuracy of numerical simulations has been improved during the past decade \cite{pdf3q,suganuma3q,jahn3q,suganuma3q-rev} that provided evidence for the Y-law at long distances. On the other hand, it is expected that at short distances the $\Delta$-law is a good approximation to the potential \cite{review,pdf3q}. However, what is still missing is a model which would incorporate the $\Delta$-law at short distances and the $Y$-law at long ones.

In this Letter we present the first example of such a model. It continues a series of studies \cite{az1,a-bar,hybrids} devoted to the static potentials in four-dimensional (pure) gauge theory by 
means of a five (ten)-dimensional effective string theory. Our reasons for continuing to pursue this model are:

(1) Because there is no string theory which is dual to QCD. It would seem very good to gain what experience we can by solving any problems that can be solved within the effective string model already at our disposal. 

(2) Because the results provided by this model are consistent with the lattice calculations and QCD phenomenology \cite{white,car,kuti}.

(3) Because analytic formulas are obtained by solving this model.

(4) Because it allows us to make predictions \cite{a-lin} which may then be tested by means of other methods, e.g., numerical simulations.

Before proceeding to the detailed analysis, let us set the basic framework. As for the quark-antiquark potential, the static $N$-quark potential can be determined from the expectation value of a Wilson loop. The loop in question, baryonic loop, is defined in a gauge-invariant manner as $W_{\nq}=\frac{1}{N!}\varepsilon_{a_1\dots a_N}\varepsilon_{a'_1\dots a'_N}\prod_{i=1}^N U^{a_ia'_i}$, with the path-ordered exponents $U^{a_ia'_i}$ along the lines shown in Figure 1. 
\begin{figure*}[ht]
\centering
\includegraphics[width=3.3cm]{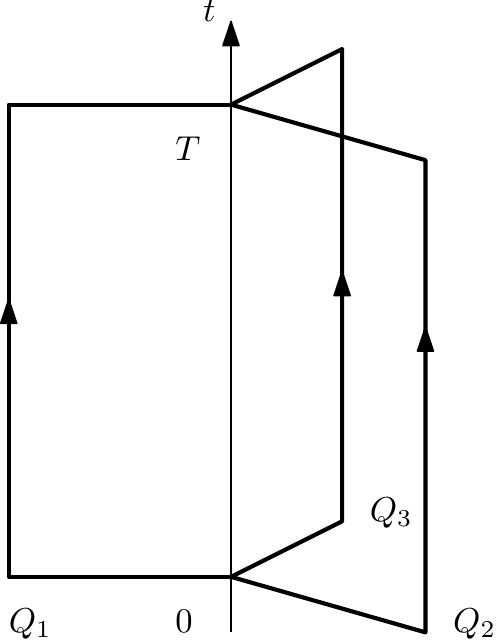}
\hspace{3.7cm}
\includegraphics[width=4.3cm]{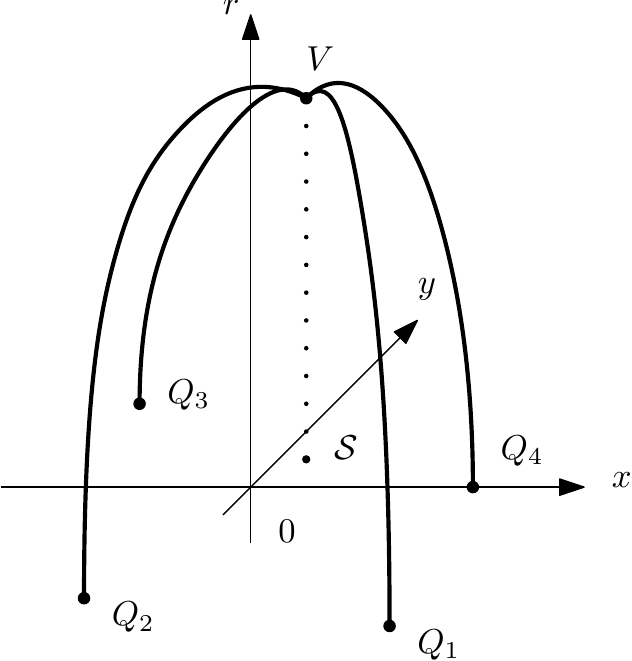}
\caption{\small{ Left: A baryonic Wilson loop in $SU(3)$ gauge theory. Right: In $SU(4)$, a configuration used to calculate the expectation value of a baryonic loop. The quarks are set on the $x\text{-}y$ plane. $V$ is a baryon vertex located at $r=r_0$ and ${\cal S}$ is its projection onto the $x\text{-}y$ plane.}}
\label{conf}
\end{figure*}
In the limit $T\rightarrow\infty$ the expectation value of the loop is simply $\langle W_{\nq}(C)\rangle\sim \ep^{-ET}$, with $E$ the ground state energy of $N$ quarks ($N$-quark potential).

In discussing baryonic Wilson loops, we adapt the formalism \cite{witten,gross} proposed within the AdS/CFT correspondence \cite{malda} to our purposes. First, we take the following ansatz for the background geometry \cite{a-pis}
\begin{equation}\label{metric10}
ds^2=\ep^{\s r^2}\frac{R^2}{r^2} \bigl(dt^2+d\vec x^2+dr^2\bigr)+\text{e}^{-\s r^2}g_{ab}^{(5)}d\omega^a d\omega^b 
\,,
\end{equation}
where $d\vec x^2=dx^2+dy^2+dz^2$. This is a deformed product of $\text{AdS}_5$ and an internal space (five-sphere) ${\bf X}$ whose coordinates are $\omega^a$. The deformation is due to the $r$-dependent warp factor, with $\s$ the deformation parameter. Such a deformation is a kind of the soft wall model of \cite{son}, where the violation of conformal symmetry is manifest in the background metric. In \eqref{metric10}, there are two free parameters to be fitted to the results of numerical simulations or quarkonia spectra. Both fits look very good \cite{white,car}.

Next, we consider the baryon vertex which is a $N$-string junction. Since we are interested in a static quark potential, we choose static gauge and then make an ansatz for the action, describing a static configuration, of the form

\begin{equation}\label{vertex}
S_{\text{vert}}=m\frac{\ep^{-2\s r^2}}{r}T\,,
\end{equation}
where $m$ and $\s$ are parameters, $r$ is independent of $t$, and $T=\int_0^T dt$. In what follows, we will assume that quarks are placed at points on the boundary of 5-dimensional deformed AdS (at $r=0$) but at the same point in the internal space. This assumption makes the problem effectively five-dimensional. Therefore the detailed structure of $\mathbf{X}$ is not important, except for the warp factor depending on the radial direction. The motivation for such a factor in \eqref{vertex} is drawn from the AdS/CFT construction, where the baryon vertex is a 5-brane \cite{witten}. Taking a term $\int dtd^5\omega\sqrt{g^{(6)}}$ from the world-volume action of the brane results in $T\ep^{-2\s r^2}/r$ if $r$ is independent of $t$. This is, of course, a heuristic argument but, as we will see, the ansatz \eqref{vertex} is quite successful: it allows us to describe the results for $N=3$ using just one parameter. 

The expectation value of the Wilson loop is schematically given by the path integral over world-sheet fields
\begin{equation}\label{ansatz}
\langle W_{\nq}(C)\rangle=\int D\Psi \ep^{-S_w}
\,,
\end{equation}
where $S_w$ is a total action of the Nambu-Goto strings and vertex. The strings are stretched between the quarks on the boundary and the baryon vertex in the interior, as sketched in Figure \ref{conf}. In principle, the integral can be evaluated approximately in terms of minimal surfaces that obey the boundary conditions. The result is written as $\langle W_{\nq}(C)\rangle=\sum_n w_n\exp[-S_n]$, where $S_n$ means a renormalized minimal area whose weight is $w_n$.

\textit{Calculating the $N$-quark potential.---} We consider a situation in which $N$ quarks are placed at the vertices of a regular $N$-sided convex polygon of side length $L$. This configuration has the symmetry group $D_N$. Hence ${\cal S}$ is a center of the polygon and all the strings have an identical profile. To compute the potential, we proceed along very similar lines to those of \cite{a-bar}. First, we take the static gauge that allows us to solve the equations of motion and determine the string profile. Next we extremize the action with respect to the location of the baryon vertex $r_0$ that results in the no-force condition at $r=r_0$. There is, however, one important distinction between the present calculation and those in the literature devoted to large $N$ gauge theories. We make an assumption that the parameter $m$ is negative. As a result, gravity pulls the vertex toward the boundary. This bends the strings and blunts the tip of the configuration \cite{a-lin}, as shown in Figure 1. 

Having found the solution, we can compute the total energy of the configuration. At the end of the day we arrive at \cite{a-lin}
\begin{widetext}
\begin{equation}\label{L}
L(\nu)=2\sin\Bigl(\frac{\pi}{N}\Bigr)\sqrt{\frac{\lambda}{\s}}
\biggl[
\int^1_0 dv\, v^2\, \ep^{\lambda(1-v^2)}\Bigl(1-v^4\ep^{2\lambda(1-v^2)}\Bigr)^{-\frac{1}{2}}+
\int^1_{\sqrt{\frac{\nu}{\lambda}}} dv\, v^2\, \ep^{\lambda(1-v^2)}\Bigl(1-v^4 \ep^{2\lambda(1-v^2)}
\Bigr)^{-\frac{1}{2}}
\biggr]
\end{equation}
and 
\begin{equation}\label{E}
E(\nu)=N\g\sqrt{\frac{\s}{\lambda}}
\biggl[
\kappa\sqrt{\frac{\lambda}{\nu}}\ep^{-2\nu}-1+
\int^1_0\,\frac{dv}{v^2}\,\biggl(\ep^{\lambda v^2}\Bigl(1-v^4\ep^{2\lambda (1-v^2)}\Bigr)^{-\frac{1}{2}}
-1\biggr)+
\int^1_{\sqrt{\frac{\nu}{\lambda}}}\,\frac{dv}{v^2}\,\ep^{\lambda v^2}\Bigl(1-v^4\,\ep^{2\lambda (1-v^2)}\Bigr)^{-\frac{1}{2}}
\biggr]+C
\,,
\end{equation}
\end{widetext}
where $\nu=\s r_0^2$, $\g=\tfrac{R^2}{2\pi\alpha'}$, $\kappa=\tfrac{m}{N\g}$, and $C$ is a normalization constant. $\lambda$ is a function of $\nu$ and $\kappa$ such that $\lambda=-\text{ProductLog}[-\nu\ep^{-\nu}(1-\kappa^2(1+4\nu)^2\ep^{-6\nu})^{-\frac{1}{2}}]$, where $\text{ProductLog}(z)$ is the principal solution for $w$ in $z=w\ep^w$ \cite{wolf}. Also note that $\nu\in[0,\nu_{\ast}]$, with $\nu_\ast$ a solution to $\nu^2=\ep^{2(\nu-1)}(1-\kappa^2(1+4\nu)^2\ep^{-6\nu})$. 

Thus, the $N$-quark potential as a function of the interquark separation is parametrically specified via the parametric functions $E(\nu)$ and $L(\nu)$. Importantly, the parameters $\g$ and $\s$ coincide with those of the quark-antiquark potential \cite{az1,hybrids} and, as a consequence, $\kappa$ is the only free parameter in the model. This is our main result.

It is worth analyzing $E(L)$ in the two limiting cases, short and long distances. In the former case we find 
\begin{equation}\label{EL-small}
E(L)=-q_{\n}\frac{\alpha_{\nq}}{L}+C+\frac{p_{\n}}{N-1}\sigma_0 L+o(L)\,,
\end{equation}
\begin{equation}\label{alpha-sigma}
\alpha_{\nq}=
-\frac{1}{q_{\n}}L_0E_0\g
\,,\quad
\sigma_0=\frac{N-1}{p_{\n} L_0}\Bigl(E_1+\frac{L_1}{L_0}E_0\Bigr)\g\,\s
\,,
\end{equation}
where $q_{\n}=\sum_{i>j}^N L/r_{ij}$, $p_{\n}=\sum_{i>j}^N r_{ij}/L$, $r_{ij}$ denotes the distance between the vertices $i$ and $j$, $L_i$ and $E_i$ are given in the Appendix. 

In the latter case, we get a generalization (star-law) of the Y-law with a single Steiner point ${\cal S}$
\begin{equation}\label{EL-large}
E(L)=\frac{N}{2\sin\bigl(\frac{\pi}{N}\bigr)}\sigma L+c+o(1)\,,
\end{equation}
with the same string tension $\sigma$ as in \cite{az1,hybrids,a-bar}
\begin{widetext}
\begin{equation}\label{sigma-c}
\sigma=\ep\g\s
\,,\quad
c=N\g\sqrt{\s}\biggl[
\frac{\kappa}{\sqrt{\nu_\ast}}\ep^{-2\nu_\ast}-1+
\int_0^1\frac{dv}{v^2}\biggl(\ep^{v^2}\Bigl(1-v^4\ep^{2(1-v^2)}\Bigr)^{\frac{1}{2}}-1\biggr)
+
\int_{\sqrt{\nu_\ast}}^1\frac{dv}{v^2}\ep^{v^2}\Bigl(1-v^4\ep^{2(1-v^2)}\Bigr)^{\frac{1}{2}}
\biggr]
+C\,.
\end{equation}
\end{widetext}
Three features of the model are worth highlighting here. First, at short distances it yields the subleading linear term \cite{zakharov}. Second, at long distances the model reduces to the star-law with the physical string tension $\sigma$, as expected \cite{a-bar}. Finally, the constant terms at short and long distances are different. Notice that $c-C$ is scheme independent and is free from divergences. This makes the model so different from the phenomenological laws.

\textit{Numerical Results and Phenomenological Prospects.---} It is of great interest to compare our model of the $N$-quark potential with the results of numerical simulations. 

Clearly, $N=3$ is of primary importance \cite{N}. In this case, we set $\g=0.176$ and $\s=0.44\,\text{GeV}^2$, i.e., to the same values as those of \cite{hybrids} used for modeling the quark-antiquark potentials of \cite{kuti}. With these parameters fixed, the model has only a single free parameter $\kappa$. We then fit it with $\alpha_{\3q}=0.125$ taken from \cite{jahn3q}. So we get $\kappa=-0.083$. The result is plotted in Figure 2, on the left. 
\begin{figure*}[ht]
\centering
\includegraphics[width=7.5cm]{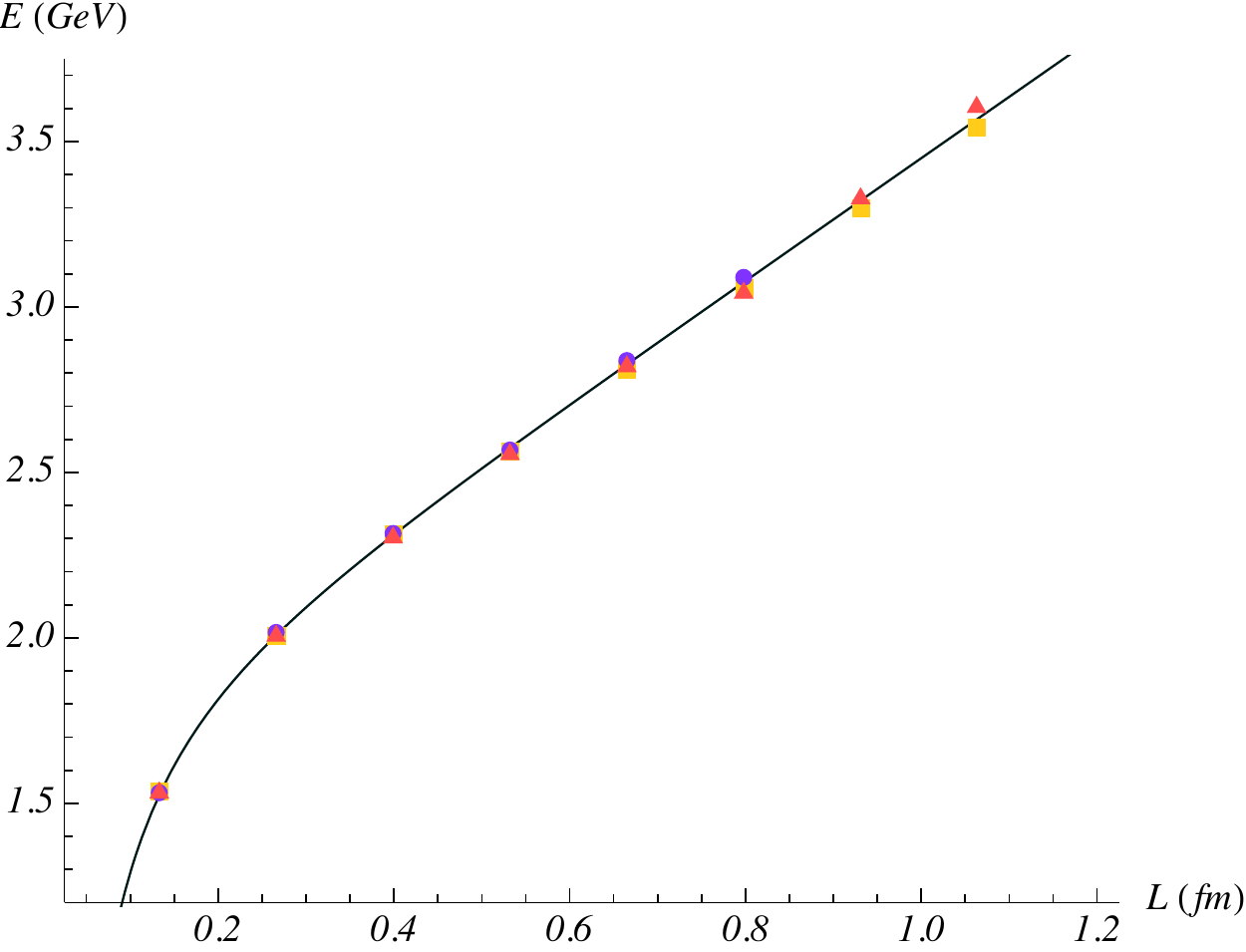}
\hfill
\includegraphics[width=7.5cm]{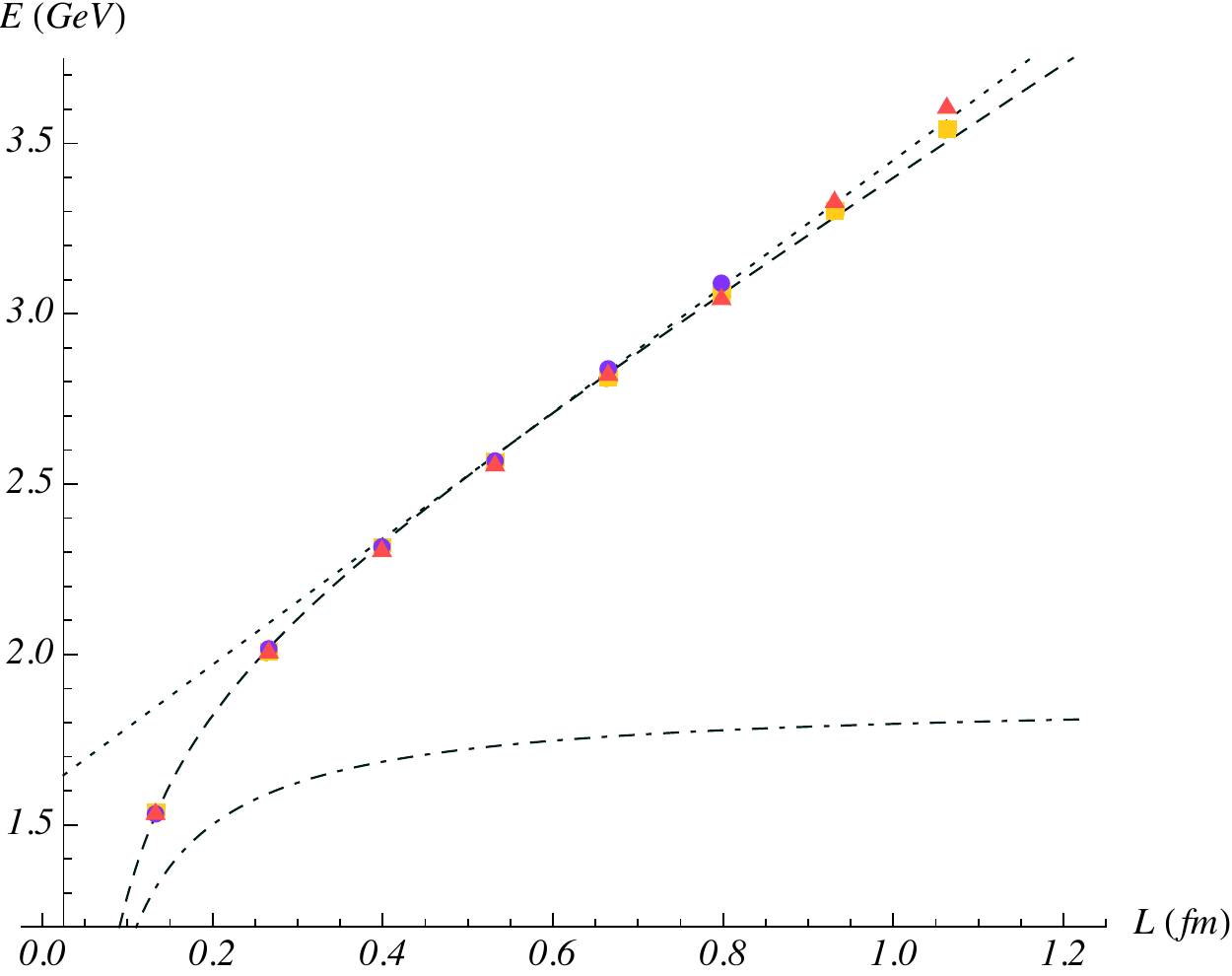}
\caption{\small{The lattice data, obtained on an equilateral triangle at $\beta=6.0$, are taken from \cite{pdf3q,jahn3q} (squares), \cite{suganuma3q} (disks), and \cite{suganuma3q-rev} (triangles). We use the normalization of \cite{jahn3q}. We don't display any error bars because they are comparable to the size of the symbols. Left: The 3-quark potential at $\g=0.176$, 
$\s=0.44\,\text{GeV}^2$, $\kappa=-0.083$, and $C=1.87\,\text{GeV}$. Right: Some asymptotic curves for the model at the same parameter values: $-3\frac{\alpha_{\3q}}{L}+C$ (dot dashed), $-3\frac{\alpha_{\3q}}{L}+C+\frac{3}{2}\sigma_0 L$ (long dashes), and $\sqrt{3}\sigma  L+c$ (short dashes).}}
\end{figure*}
We see that the model reproduces the lattice data remarkably well. Also note that a fit to $\alpha_{\3q}=0.129$ of \cite{suganuma3q-rev} doesn't change the picture as the discrepancy between the data of \cite{jahn3q} and \cite{suganuma3q-rev} is negligible at small distances. 

For completeness, it is worth making a couple of estimates. First, from \eqref{alpha-sigma} we get $\alpha_{\3q}/\alpha_{\qqb}\approx 0.495$, where $\alpha_{\qqb}=(2\pi)^3\Gamma^{-4}\bigl(\tfrac{1}{4}\bigr)\g$ \cite{az1,malda2}. Thus, the relation between the 
"Coulomb" coefficients found in perturbative QCD holds with good accuracy in our model. 
This looks puzzling as we consider small distances but not very small ones of perturbative QCD. Second, from \eqref{alpha-sigma} and \eqref{sigma-c}, we find $\sigma_0/\sigma\approx 1.007$ that favours the $\Delta$-law at short distances, as also noted in \cite{jahn3q}.

For practical purposes, the parametric form of the potential looks somewhat awkward. It is instructive to compare the lattice data to the asymptotic behavior of $E(L)$ to see what happens. In Figure 2 we have plotted the results. As can be seen, in the range of interest a single Coulomb-type term doesn't yield a satisfactory description. But if one adds an additional linear term, then the situation will improve. Such a two-component model, almost the $\Delta$-law, does describe the data quite well in the range $0.1\,\text{fm}\leq L\lesssim 0.6\,\text{fm}$. However, at longer distances $0.6\,\text{fm}\lesssim L\leq 1.2\,\text{fm}$, it becomes less accurate than a single linear term from the Y-law, as already noted in \cite{suganuma3q}. Thus the model we propose smoothly interpolates between the $\Delta$ and Y-laws with a transition at $L\approx 0.6\,\text{fm}$.

Unfortunately, in lattice gauge theory very little is known about the $N$-quark potential if $N>3$. Even in SU(4) \cite{pdf3q}, the availability of data is much more limited than it should be to make a consistent comparison with our predictions.

In the physically interesting case $N=3$, the model incorporates pairwise interactions, the $\Delta$-law, at short distances and a genuine three-body interaction at longer ones. Note that one can think of the pairwise interaction 
as that of a quark and a diquark thanks to a string stretched in between. Such an interaction occurs at intermediate scales in between very small scales, where quarks are asymptotically free, and large scales, where quarks are strongly coupled.

\textit{Summary and discussion.---} In this Letter, we have modeled the $N$-quark potential using the now standard ideas motivated by gauge-string duality. Our work based on the background geometry \eqref{metric10}, which is singled out by the earlier works \cite{hybrids,a-pis}, provides the first convincing example of interpolation between the $\Delta$ and $Y$-laws. Mathematically, the potential is described by a complicated function whose asymptotic behavior is given by the $\Delta$ and $Y$-laws. 

The model we are developing is an effective string theory based on the Nambu-Goto formalism in a curved space. Therefore it has some limitations including: the issue of a L\"uscher-like correction on a curved background and the issue of attraction between the baryon vertex and boundary. What could be the reason for the latter? $m$ is a result of a resummation of infinitely many terms ($\alpha'$ corrections) in the five-brane action. Is it negative because the brane tension is negative  \cite{negativeT}, and if so, does it lead to instability? These questions have no obvious or immediate answers. Hopefully, it will be resolved in the future by using the Green-Schwarz formalism, already developed for strings on $\text{AdS}_5\times\mathbf{S}^5$ \cite{rr}. Obviously, finding the way to the string description of QCD is a challenging and difficult problem. In the meantime, lattice gauge theory and effective string models will remain the main tools of investigation.

\textit{Acknowledgments.---} We would like to thank P. de Forcrand, A. Leonidov, and P. Weisz for stimulating discussions, and P. de Forcrand and H. Suganuma for providing us with the results of lattice simulations. This work was supported in part by the Alexander von Humboldt Foundation. Finally we thank the Arnold Sommerfeld Center for Theoretical Physics and CERN Theory Division for the warm hospitality. 

\appendix
\renewcommand{\theequation}{A.\arabic{equation}}
\setcounter{equation}{0}
\textit{Appendix: Some useful formulas.---} In studying the short distance behavior of $E(L)$, the following two facts are useful. First, $L$ goes to zero as $\nu\rightarrow 0$. Second, for small $\nu$ and $\lambda$ we have $\nu(\lambda)=\rho^{\frac{1}{2}}\lambda+\bigl(1-2\kappa^2-\rho^{\frac{1}{2}}\bigr)\lambda^2+o(\lambda^2)$, with $\rho=1-\kappa^2$.

Expanding \eqref{L} and \eqref{E} to subleading order in $\lambda$, we find 
\begin{equation}\label{le}
L=\sqrt{\frac{\lambda}{\s}}\bigl(L_0+L_1\lambda\bigr)\,,\,\,
E=\g\sqrt{\frac{\s}{\lambda}}\bigl(E_0+E_1\lambda\bigr)+C\,,
\end{equation}
together with 
\begin{gather*}
L_0=\frac{1}{2}\sin\bigl(\tfrac{\pi}{N}\bigr)
{\cal B}\bigl(\kappa^2;\tfrac{1}{2},\tfrac{3}{4}\bigr)
\,,\\
L_1=\frac{1}{2}\sin\bigl(\tfrac{\pi}{N}\bigr)
\Bigl({\cal B}\bigl(\kappa^2;-\tfrac{1}{2},\tfrac{3}{4}\bigr)-
{\cal B}\bigl(\kappa^2;-\tfrac{1}{2},\tfrac{5}{4}\bigr)-\eta\Bigr)
\,,\\
E_0=N
\Bigl(\kappa\rho^{-\frac{1}{4}}+
\frac{1}{4}{\cal B}\bigl(\kappa^2;\tfrac{1}{2},-\tfrac{1}{4}\bigr)\Bigr)
\,,\\
E_1=N\Bigl(
-2\kappa\rho^{\frac{1}{4}}+
\frac{1}{4}{\cal B}\bigl(\kappa^2;\tfrac{1}{2},\tfrac{1}{4}\bigr)+
\frac{1}{2}\frac{L_1}{\sin\bigl(\tfrac{\pi}{N}\bigr)}
\Bigr)
\,.
\end{gather*}
Here $\eta=2\rho^{\frac{1}{4}}\vert\kappa\vert^{-1}\bigl(1-2\kappa^2-\rho^{\frac{1}{2}}\bigr)$, ${\cal B}(z;a,b)=B(a,b)+B(z;a,b)$, and $B(z;a,b)$ is the incomplete beta function. Then a simple algebra leads to the explicit formulas for $\alpha_{\nq}$ and $\sigma_0$.


\end{document}